\documentclass[12pt]{article}

\usepackage[a4paper,left=30mm,right=20mm,top=20mm,bottom=20mm]{geometry}
\usepackage{graphics}
\usepackage{amsmath}
\usepackage{amssymb}
\usepackage{epsfig}
\usepackage{cite}
\usepackage{xcolor}
\usepackage[unicode]{hyperref}
\newcommand{\bea}{\begin{eqnarray}}
\newcommand{\eea}{\end{eqnarray}}
\newcommand{\be}{\begin{equation}}
\newcommand{\ee}{\end{equation}}

\title{A simple model for nuclear modification of parton distribution functions}
\author{A.V.~Kotikov$^{1}$, A.V.~Lipatov$^{2}$}
\begin{document}

\maketitle

\begin{center}
%{\it $^{1}$School of Physics and Astronomy, Sun Yat-sen University, 519082 Zhuhai, China}\\
{\it $^{1}$Joint Institute for Nuclear Research, 141980 Dubna, Moscow region, Russia}\\
{\it $^{2}$Skobeltsyn Institute of Nuclear Physics, Lomonosov Moscow State University, 119991 Moscow, Russia}

\end{center}

\vspace{0.5cm}

\begin{center}

{\bf Abstract }

\end{center}

\indent

A model for nuclear medium modification of parton densities is presented.
%\textcolor{blue}{
%We review QCD motivated nuclear modifications of parton densities.
%We confront theoretical approaches to a global analysis of all the%}
The approach is based on the  global analysis of
available deep inelastic scattering data for different nuclear targets, %\textcolor{blue}{
%with an emphasis on%}
within  the rescaling model combined with taking into account
  the effects of Fermi motion. %\textcolor{blue}{
%  Here the%}
  The
  scale dependence is implemented into the
DGLAP-evolved quark and gluon densities
in a proton derived analytically at the leading order of QCD coupling.
By fitting the rescaling parameters to experimental data on the ratio
$F_2^A(x,Q^2)/F_2^{A^\prime}(x,Q^2)$ for several nuclear targets $A$ and $A^\prime$, we obtain predictions
for nuclear parton distributions,
even for unmeasured nuclei. The effects of nuclear modifications are investigated
with respect to the mass number $A$.
%\textcolor{blue}{
We %}
%Our results
highlight distinct shadowing and antishadowing behaviors for gluons and quarks.

\vspace{1.0cm}

\noindent{\it Keywords:} deep inelastic scattering, parton densities in a proton and nuclei, EMC effect.

%\newpage
\vspace{1.0cm}

\newpage

\section{Introduction} \indent

The study of deep inelastic scattering (DIS) of leptons on nuclei shows significant
effects of nucleon interactions within the nucleus, challenging
a naive picture of the nucleus as a system of quasi-free nucleons (see reviews\cite{NuclReview-1, NuclReview-2, NuclReview-3, MKla24}).
Nuclear medium effects on parton (quark and gluon) distribution functions (PDFs) meet a lot of interest from both
experimental and theoretical
points\cite{EMC-CCuSn, EMC-CCa1, EMC-CCa2, EMC-Cu, NMC-CCaovLi, NMC-HeCCaCCaovLi, NMC-LiC, NMC-BeAlCaFeSnPbovC, NMC-SnovC, SLAC-Fe, BCDMS-NFe, BCDMS-Fe, E665-CCaPb, E665-Xe, JLab-HeCBe, CLAS-CAlFePb}.
In particular, a detailed understanding of nuclear modifications of the parton densities (nPDFs) is a fundamental
problem of nuclear and high energy physics at present and it is crucial for any theoretical description of $pA$ and $AA$ collisions
at modern (LHC, RHIC) and future colliders (FCC-he, EiC, EicC, CEPC, NICA).
Usually, the nuclear modification factor is
defined as a ratio of per-nucleon DIS structure functions in nuclei $A$ and
deuteron\footnote{Assuming that the nuclear effects in deutron are negligible, so the deuteron is considered as a system of a free proton and neutron.
  We note, however, that studies of nuclear effects in the deuteron can be found\cite{SAlekhin22, SAlekhin17, nIMP, nNNPDF3}. See also discussions
    \cite{SKulagin17}.} $D$, $R^A = F_2^A(x,Q^2)/F_2^D(x,Q^2)$, or rather ratio of corresponding parton densities,
is introduced to investigate the behavior across different kinematic
regions:
shadowing ($x \leq 0.1$), anti-shadowing ($0.1 \leq x \leq 0.3$), valence quark dominance
($0.3 \leq x \leq 0.7$), and Fermi motion ($x \geq 0.7$).
The shadowing and anti-shadowing effects refer to $R^A < 1$ and $R^A > 1$, whereas the EMC effect\cite{EMCEffect} and Fermi
motion refer to the slope of $R^A$ in the valence-dominant region and rising of $R^A$ at larger $x$.
%\footnote{
%\textcolor{blue}{
Note that the nuclear medium effect was first discovered\cite{EMCEffect} by the European Muon Collaboration (EMC)
	in the domain of valence quark dominance. The investigations of shadowing and antishadowing (see\cite{Stodolsky, NiZa75, ColorDipoleModel-3})
	had started before the availability of the EMC experimental data\cite{EMC-CCa1, EMC-CCa2}
        (see also\cite{Ni81}  for overview). Moreover, antishadowing as a phenomenon was introduced in Ref.\cite{NiZa75}.%}

Unfortunately, at present there is no commonly recognized framework to describe nuclear PDF modifications
in the whole kinematical range of $x$, as well as corresponding nuclear dependence.
Two main approaches exist: global fits to nuclear data using some empirical starting nPDFs
followed by the standard Dokshitzer-Gribov-Lipatov-Altarelli-Parisi
(DGLAP) equations\cite{DGLAP} to describe their scale dependence (see recent studies\cite{nCTEQ, nNNPDF3, EPPS21, nIMP, PDuwen22, HKhan21, IHelen22}
  and references therein)
and model-based approaches (see, for example,\cite{SKulagin17, RWang18} and discussions therein).

The first technique is in a close analogy with the standard derivation of the PDFs (see review\cite{MKla24}).
Models of nuclear medium modifications can be roughly divided into two categories: models based on conventional nuclear physics and
  models inspired by QCD (see\cite{RWang18}). The first of them %Models based on conventional nuclear physics
  typically take into account the decrease in the nucleon mass in the medium,
  leading to so-called $x$-rescaling models\cite{CGarciz84, MStas84, SAAkulin85, LLFran87, HJung88, CCiofi99} and off-shellness
  corrections\cite{GDunne86, FGross92, SKulagin94, KulaginPetti-1, KulaginPetti-2}. The QCD-inspired models typically require an
  increase in quark confinement or a simple increase
  in the nucleon radius (nucleon swelling  \cite{RWang18, nIMP}). A larger nucleon corresponds to a higher probe resolution. In the language of
  QCD evolution, $Q^2$-rescaling\cite{RescalingModel-Apps1, AKotikSZ17, RescalingModel-Apps2, RescalingModel-1, FClose88} is used to interpret the effect.

Of course, all these analyses\cite{nCTEQ, nNNPDF3, EPPS21, nIMP, KulaginPetti-1, KulaginPetti-2, RescalingModel-1, RescalingModel-2} show remarkable progress in the last decade.
Nevertheless, quark and especially gluon
densities in nuclei still have large uncertainties in the whole $x$ region due to shortage
of the experimental data and/or limited kinematic coverage of the latter\cite{nPDFs-unc-1, nPDFs-unc-2, nPDFs-unc-3}.

In the framework of the $Q^2$-rescaling model (everywhere below --- rescaling model),
 the nuclear medium modification was investigated\cite{RescalingModel-1, RescalingModel-Apps1, RescalingModel-Apps2}.
Initially, this model was proposed for the  valence-dominant region and it is
based on a suggestion that the effective confinement size of gluons and
quarks in the nucleus is greater than in the free nucleon.
Such shift in the confinement scale results in relation
between the PDFs and nPDFs through a simple rescaling of their arguments\cite{RescalingModel-1, RescalingModel-2}.
So, the rescaling model demonstrates the features inherent in both approaches: there is a connection between
PDFs and nPDFs arising due to the shift in the scale and, at the same time, both PDFs and nPDFs obey the DGLAP equations.
Relatively recently the rescaling model has been extended to a low $x$ range\cite{RescalingModel-Apps1}
%\textcolor{blue}{
(Note that in the framework of the
  $x$-rescaling model, the low $x$ range was firstly investigated\cite{Ni93}).%}.
It was shown\cite{RescalingModel-Apps1, AbKoLi23, RescalingModel-Apps2} that good agreement with available experimental data on
the $F_2^A(x,Q^2)/F_2^D(x,Q^2)$ ratios
at low and moderate $x$ could be achieved by fitting the corresponding rescaling parameters.

The aim of this work is to extend and
improve earlier considerations\cite{RescalingModel-Apps1, RescalingModel-Apps2}.
Firstly, we employ updated analytical expressions for PDFs
obtained very recently\cite{KL2025} from the solution of the DGLAP evolution equations
at the leading order (LO) of QCD coupling.
These expressions rely on exact asymptotics at small and large $x$ and contain subasymptotic terms that are
fixed by the momentum conservation and/or Gross-Llewellyn-Smith and Gottfried sum rules.
Some phenomenological parameters have been determined from a rigorous fit to precision
BCDMS, H1 and ZEUS experimental data on the proton structure function $F_2(x,Q^2)$ in a wide
kinematical region, $2 \cdot 10^{-5} \leq x \leq 0.75$ and $1.2 \leq Q^2 \leq 30000$~GeV$^2$.
This is in contrast with the previous studies\cite{RescalingModel-Apps1, RescalingModel-Apps2}, where only low-$x$ asymptotics were
considered.
The second improvement is related with the taking into account effects of the Fermi motion.
Of course, the latter is necessary to provide a consistent description of nuclear modifications
across the entire kinematic range.
Thirdly, to extract the rescaling parameters we perform a
global fit to available experimental data on the ratios
$F_2^A(x,Q^2)/F_2^{A^\prime}(x,Q^2)$ for different nuclear targets $A$ and $A^\prime$
collected by the EMC\cite{EMC-CCuSn, EMC-CCa1, EMC-CCa2, EMC-Cu}, NMC\cite{NMC-CCaovLi, NMC-HeCCaCCaovLi, NMC-LiC, NMC-BeAlCaFeSnPbovC, NMC-SnovC},
SLAC\cite{SLAC-Fe}, BCDMS\cite{BCDMS-NFe, BCDMS-Fe}, E665\cite{E665-CCaPb, E665-Xe}, JLab\cite{JLab-HeCBe} and CLAS\cite{CLAS-CAlFePb} Collaborations.
Then we investigate the dependence of our
results on the mass number $A$,
derive predictions for corresponding nuclear parton distributions
in simple analytical form and investigate
effects of nuclear modifications.
In a sense, our present study is a continuation of previous investigations\cite{RescalingModel-Apps1, RescalingModel-Apps2, KL2025}.

The outline of our paper is following. In Section~2 we
list small-$x$ and large-$x$ asymptotics which are used to construct the PDFs in
the whole kinematical region. In Section~3 we recall basic formulas of
rescaling model and extend it with Fermi motion.
Section~4 presents our numerical results and discussions. Section 5 contains our conclusions.

\section{Structure function $F_2(x,Q^2)$ and parton densities in a proton} \indent

We start from some basic formulas used in our calculations.
It is well known that
the proton structure function $F_2(x,Q^2)$ at the leading order (LO) of QCD coupling can be expressed as:
\begin{gather}
  F_2(x,Q^2) = \sum_{i=1}^{N_f} e_i^2 \left[f_{q_i}(x,Q^2) + f_{\bar{q}_i}(x,Q^2) \right],
\label{S1.14}
\end{gather}
\noindent
where $e_i$ is the fractional electric charge of quark $q_i$, $N_f$ is the number of active
quark flavors, $f_{q_i}(x,Q^2)$ and $f_{\bar q_i}(x,Q^2)$ represent the quark and antiquark densities in a proton (multiplied by $x$), respectively.
In the fixed-flavor-number-sheme (FFNS) with $N_f = 4$, where $b$ and $t$ quarks are separated out, we have:
\begin{gather}
  F_2(x,Q^2) = \frac{5}{18} \, f_{SI}(x,Q^2) + \frac{1}{6} f_{NS}(x,Q^2),
\label{S1.15}
\end{gather}
\noindent
where the singlet part $f_{SI}(x,Q^2)$ contains the valence and sea quark parts:
\begin{gather}
	f_{V}(x,Q^2) = f_u^V(x,Q^2) + f_d^V(x,Q^2), \quad f_{S}(x,Q^2) = \sum_{i=1}^{4} \left[f_{q_i}^S(x,Q^2) + f_{\bar{q}_i}^S(x,Q^2) \right], \nonumber \\
	f_{SI}(x,Q^2) = \sum_{i=1}^{4} \left[f_{q_i}(x,Q^2) + f_{\bar{q}_i}(x,Q^2) \right] = f_{V}(x,Q^2) + f_{S}(x,Q^2).
	\label{eq-FSI}
\end{gather}
\noindent
The nonsinglet part $f_{NS}(x,Q^2)$ contains the difference between up and down quarks:
\begin{gather}
	f_{NS}(x,Q^2) = \sum_{q=u,\,c} \left[f_{q}(x,Q^2) + f_{\bar{q}}(x,Q^2) \right] - \sum_{q=d,\,s} \left[f_{q}(x,Q^2) + f_{\bar{q}}(x,Q^2) \right].
	\label{eq-FNS-NF4}
\end{gather}
\noindent
The non-singlet and valence parts of quark distribution functions can be represented in the following form\cite{KL2025, PDFs-our} (see also\cite{PDFs-our-previous, PDFs-early-approach}):
\begin{gather}
	f_i(x,\mu^2) = \left[A_i(s)x^{\lambda_i}(1 -x) + \frac{B_i(s)\, x}{\Gamma(1+\nu_i(s))} + D_i(s)x (1 -x) \right] (1-x)^{\nu_i(s)},
	\label{eq1-NSV}
\end{gather}
\noindent
where $i = NS$ or $V$, $s = \ln \left[\alpha_s(Q_0^2)/\alpha_s(Q^2)\right]$ and
\begin{gather}
	A_{i}(s)=A_{i}(0) e^{-d(n_i)s}, \quad B_{i}(s) = B_{i}(0) e^{-p s}, \quad \nu_{i}(s)=\nu_{i}(0)+r s, \nonumber \\
	r = \frac{16}{3\beta_0}, \quad p=r\left(\gamma_{\rm E}+\hat{c}\right), \quad \hat{c}=-\frac{3}{4}, \quad d(n)=\frac{\gamma_{NS}(n)}{2\beta_0}, \quad n_i = 1 - \lambda_i.
	\label{eq2-NSV}
\end{gather}
\noindent
Here $\Gamma(z)$ is the Riemann's $\Gamma$-function, $\gamma_{\rm E} \simeq 0.5772$ is the Euler's constant,
$\beta_0 = 11 - 2N_f/3$ is the LO QCD $\beta$-function, $\lambda_{NS} = \lambda_V = 0.5$, $\gamma_{NS}(n)$ is the LO non-singlet anomalous
dimension and $A_i(0)$, $B_i(0)$ and $\nu_i(0)$ are the free parameters.
Note that the expression~(\ref{eq1-NSV}) is constructed as a combination of
the small-$x$ part proportional to $A_i(s)$, large-$x$ asymptotic proportional to $B_i(s)$ and an
additional term proportional to $D_i(s)$. The latter is subasymptotics in both these regions
and its scale dependence is fixed by the Gross-Llewellyn-Smith and Gottfreed sum rules\cite{KL2025, PDFs-our}:
\begin{gather}
	D_i(s) = \left(2 + \nu_i(s) \right) \left[N_i - A_i(s) {\Gamma(\lambda_i) \Gamma(2 + \nu_i(s)) \over \Gamma(\lambda_i + 2 + \nu_i(s))} - {B_i(s) \over \Gamma(2 + \nu_i(s))} \right],
	\label{eq3-NSV}
\end{gather}
\noindent
%\textcolor{blue}{
where $N_V = 3$ and\cite{NMCGott} (see\cite{KL2025} and references therein for more information)
\be
N_{NS} \equiv I_G(Q^2) = 0.705 \pm 0.078\,.
\label{Gottfried}
\ee

%\textcolor{blue}{
  Basing on\cite{KL2025} (see also\cite{KoKriSha18}), we use the result (\ref{Gottfried}) for $I_G(Q^2)$, which is very different from the theoretical
  prediction $I_G(Q^2)=1$ based on the quark sea symmetry. In principle, such a violation could occur not for partons inside a free nucleon, but under nuclear medium
  modification. We plan to explore this possibility in our future studies.
  %}

% (see\cite{KL2025} and references therein for more information).
The singlet part of quark densities and gluon distribution
in a proton can be represented as combinations of "$\pm$" terms\cite{KL2025, PDFs-our} (see also\cite{PDFs-our-previous, RescalingModel-Apps1}):
\begin{gather}
	f_i(x,Q^2) = f_i^+(x,Q^2) + f_i^-(x,Q^2),
	\label{eq1-Sg}
\end{gather}
\noindent
where $i = SI$ or $g$ and
%\textcolor{blue}{
(note that basing on\cite{gDAS1, gDAS2, gDAS3, Rujula74}, we exploit the non-Regge behavior for singlet quarks and gluons at small $x$.
    The corresponding parametrizations with Regge behavior at small $x$ \cite{LoYndu1981, NNikolaev94} can be found\cite{LoYndu1984}.)%}
\begin{gather}
	f_{SI}^+(x,Q^2) = \Bigg[ {N_f \over 9} \left(A_{g} + {4\over 9}A_q\right) \rho I_1(\sigma) e^{- \bar d^{+} s} (1-x)^{m_{q}^+} + D^+(s) \sqrt x (1 - x)^{n^+} - \nonumber \\
	- {K^+ \over \Gamma(2 + \nu^+(s))} \times {B^+(s)x \over \hat c - \ln(1 - x) + \Psi(2 + \nu^+(s))} \Bigg] (1 - x)^{\nu^{+}(s) + 1},
	\label{eq-fSp}
\end{gather}
\begin{gather}
	f_{SI}^-(x,Q^2) = \Bigg[ A_{q}e^{- d^{-} s} (1-x)^{m_{q}^-} + {B^-(s)x \over \Gamma(1 + \nu^-(s))} + \nonumber \\
	+ D^{-}(s) \sqrt x (1 -x)^{n^-} \Bigg] (1 - x)^{\nu^{-}(s)},
	\label{eq-fSm}
\end{gather}
\begin{gather}
	f_{g}^+(x,Q^2) = \Bigg[ \left(A_{g} + {4\over 9}A_q\right) I_0(\sigma) e^{- \bar d^{+} s}(1 - x)^{m_{g}^+} + {B^+(s) x\over \Gamma(1 + \nu^+(s))} \Bigg] (1 - x)^{\nu^{+}(s)},
	\label{eq-fgp}
\end{gather}
\begin{gather}
	f_{g}^-(x,Q^2) = \Bigg[- {4\over 9} A_{q} e^{- d^{-} s} (1 - x)^{m_g^-} + \\ \nonumber
	+ {K^- \over \Gamma(2 + \nu^-(s))} \times {B^-(s)x \over \hat c - \ln(1 - x) + \Psi(2 + \nu^-(s))}\Bigg](1 - x)^{\nu^{-}(s)+1}.
	\label{eq-fgmtld11}
\end{gather}
\noindent
Here $\Psi(z)$ is the Riemann's $\Psi$-function, $I_0(z)$ and $I_1(z)$ are the modified Bessel functions and
\begin{gather}
		\label{eq-parameters1}
	\nu^\pm(s) = \nu^\pm(0) + r^\pm s, \quad B^\pm(s) = B^\pm(0) e^{-p^\pm s}, \quad p^\pm = r^\pm (\gamma_{\rm E} + \hat c^\pm), \nonumber \\
	r^+ = {12\over \beta_0}, \quad r^- = {16\over 3 \beta_0}, \quad \hat c^+ = - {\beta_0 \over 12}, \quad \hat c^- = - {3\over 4}, \quad K^+ = {3N_f \over 10}, \quad K^- = {2\over 5}, \nonumber \\
	\rho = {\sigma \over 2 \ln(1/x)}, \quad \sigma = 2 \sqrt{|\hat d^+|s \ln{1\over x}}, \\ \nonumber
	\hat d^+ = - {12\over \beta_0}, \quad \bar d^+ = 1 + {20 N_f\over 27\beta_0}, \quad d^- = {16 N_f \over 27 \beta_0}
\end{gather}
\noindent
with $A_g$, $A_q$, $B^\pm(0)$, $\nu^\pm(0)$, $m_q^\pm$, $m_g^\pm$ and $n^\pm$ being the free parameters.
The expressions for subasymptotic terms $D^\pm(s)$ were derived from the momentum conservation law
and could be found elsewhere\cite{KL2025}. Note that LO small-$x$ asymptotics in the formulas above were obtained in the so-called generalized
doubled asymptotic scaling (DAS) approximation\cite{gDAS1, gDAS2, gDAS3}. In this approximation, flat initial conditions, $f_g(x,Q^2_0) = A_g$
and $f_S(x,Q^2_0) = A_q$, can be used.
All the phenomenological parameters involved in~(5) --- (13) were determined\cite{KL2025} by fitting the precise data on
proton structure function $F_2(x,Q^2)$.

These expressions will be used below as one of the cornerstones
in the analysis of experimental data on nuclear structure function ratios $F_2^A(x, Q^2)/F_2^{A^\prime}(x,Q^2)$
for different nuclear targets within the framework of the rescaling model.

\section{Model of nuclear modifications} \indent

As a model of nuclear modifications of listed above parton densities, we consider the combination of
the rescaling model\cite{RescalingModel-1} and Fermi motion.
So, there is simple relationship between ordinary PDFs and nPDFs via a shift in the kinematical variable $Q^2$
proposed within the rescaling model (see also\cite{RescalingModel-2}).
For a nucleus $A$, the valence and nonsinglet parts are modified as:
\begin{gather}
  f_{i}^A(x,Q^2) = f_{i}(x,Q^2_{A,\,i}),
  \label{va.1a}
\end{gather}
\noindent
where $i = V$ or $NS$ and scale $Q^2_{A,\,i}$ is related to $Q^2$ by:
\begin{gather}
	s^A_i \equiv \ln \left(\frac{\ln\left(Q^2_{A,\,i}/\Lambda^2_{\rm QCD}\right)}{\ln\left(Q^2_{0}/\Lambda^2_{\rm QCD}\right)}\right) = s +\ln \left(1+\delta^A_i \right),
	\label{sA}
\end{gather}
\noindent
with $\delta^A_i$ being the scale independent free parameters (see\cite{RescalingModel-Apps1} and references therein)
and analytical expressions for $f_i(x,Q^2)$ are given by~(\ref{eq1-NSV}) and~(\ref{eq2-NSV}).

Exploiting the fact that rise of parton densities increases with increasing $Q^2$, the
rescaling model was extended\cite{RescalingModel-Apps1} to low $x$, or shadowing region.
Actually, if we change the scale in the QCD evolution to be less than $Q^2$,
one can immediately reproduce the nuclear shadowing effects observed in the global fits.
At low $x$, the singlet part of quark densities and gluon distribution in a proton
are non-negligible only and each of them has two ("$+$" and "$-$") independent components.
Therefore, one has two additional free parameters to be fit from the nuclear data.
Thus,
\begin{gather}
  f_i^A(x,Q^2) = f_i^{A,\,+}(x,Q^2) + f_i^{A,\,-}(x,Q^2), \quad f_i^{A,\,\pm}(x,Q^2) = f_i^{\pm}(x,Q^2_{A,\,\pm}),
\end{gather}
\noindent
where $i = SI$ or $g$ and expressions for $f_i^\pm(x,Q^2)$ are given by~(9) --- (12).
The definition of $Q^2_{A,\,\pm}$ is the same as above and
corresponding values of $s^A_\pm$ turned out to be\cite{RescalingModel-Apps1}
\begin{gather}
 s^A_\pm \equiv  \ln \left(\frac{\ln\left(Q^2_{A,\,\pm}/\Lambda^2_{\rm QCD}\right)}{\ln\left(Q^2_{0}/\Lambda^2_{\rm QCD}\right)}\right) = s +\ln \left(1+\delta^A_\pm \right).
\end{gather}
\noindent
where the free parameters $\delta^A_\pm$ are scale independent, have to be negative and could be determined from the data\cite{RescalingModel-Apps1,RescalingModel-Apps2}.

Another improvement that was made in this work in comparison with early studies\cite{RescalingModel-Apps1,RescalingModel-Apps2}
is taking into account the Fermi motion of nucleon inside the  nuclear target.
As it was already mentioned above, the Fermi smearing deforms the nuclear structure
function mainly at large $x > 0.7$.
For nonsinglet and valence parts, such deformation could be described by the convolution\cite{FermiMotion1, FermiMotion2}
\begin{gather}
	f_i^{A(F)}(x,Q^2)=\frac{1}{R_{NS}} \int\limits^{y_{\rm max}}_{y_{\rm min}}\frac{dy}{y} \left[yf_N(y)\right] f_i^A\left(x/y,Q^2\right),
	\label{Fermi1}
\end{gather}
\noindent
where $i = V$ or $NS$ and
\begin{gather}
  f_N(y)=(\eta_+ - y)(y - \eta_-), \quad \eta_{\pm} = 1 - {B_A\over m_N} \pm s_F, \quad R_{NS}=\frac{4}{3}s_F^3, \quad s_F = {k_F \over m_N}.
  	\label{Fermi2}
\end{gather}
\noindent
Here $m_N$ is the nucleon mass,
$k_F \simeq 200$~MeV is the average nucleon Fermi momentum for the nuclei
and $B_A$ is the nuclear binding energy per nucleon.
The integration limits are $y_{\rm min} = \max(x, 1 - B_A/m_N - s_F)$ and $y_{\rm max} = \min(A, 1 - B_A/m_N + s_F)$.
Using~(\ref{Fermi1}) and~(\ref{Fermi2}), we have
\begin{gather}
  \int\limits^1_0 \frac{dx}{x} f^{A(F)}_i(x,Q^2)=\int\limits^1_0 \frac{dx}{x} f_i^A(x,Q^2) = N_i.
\end{gather}
\noindent
Thus, Fermi motion keeps the Gross-Llewellyn-Smith and Gottfreed sum rules.
Similarly, for the singlet and gluon parts we have
\begin{gather}
	f_i^{A(F)}(x,Q^2)=\frac{1}{R_{SI}} \int\limits^{y_{\rm max}}_{y_{\rm min}}\frac{dy}{y}\left[yf_N(y)\right] f_i^A\left(x/y,Q^2\right),
	\label{Fermi3}
\end{gather}
\noindent
where $i = SI$ or $g$ and
\begin{gather}
	  \label{Fermi4}
  R_{SI}=\frac{4}{3}s_F^3 \eta.
\end{gather}
\noindent
One can easily obtain that
\begin{gather}
	\int\limits^1_0 dx \left[f^{A(F)}_{SI}(x,Q^2)+f^{A(F)}_{g}(x,Q^2)\right]=\int\limits^1_0 dx \left[f_{SI}^A(x,Q^2)+f_{g}^A(x,Q^2)\right] = 1,
	\label{MCF}
\end{gather}
therefore, taking into account the Fermi motion keeps the momentum conservation law.
The nuclear binding energy $B_A$ is precisely measured\cite{BindingEnergy}.
Below we use the proposed empirical formula\cite{nIMP}
to describe the nuclear dependence of the $k_F$:
\begin{gather}
  k_F(Z,A) = k_F^p \left(1 - A^{-t_1}\right) {Z\over A} + k_F^n\left(1 - A^{-t_2}\right){A - Z \over A},
  \label{Fermi5}
\end{gather}
\noindent
where $k_F^p = 365$~MeV, $k_F^n = 231$~MeV, $t_1 = 0.479$, $t_2 = 0.528$
and $Z$ and $A$ are the proton number and mass number, respectively.

Thus, the full model that was described above combines rescaling and Fermi motion effects.
The expressions~(\ref{va.1a}) --- (\ref{Fermi4}) and~(\ref{Fermi5}), being accompanied with analytical formulas~(\ref{eq1-NSV}) --- (13)
and corresponding rescaling parameters $\delta^A_V$, $\delta^A_{NS}$ and $\delta^A_\pm$
provide one with a possibility to evaluate nPDFs for any nuclear target $A$.
The nuclear structure functions $F_2^A(x, Q^2)$ as well as their ratios
for various nuclei can be calculated in the same way as it was done for the proton, according to general formula~(\ref{S1.15}).
The values of rescaling parameters are determined below.

\section{Numerical results} \indent

To extract the rescaling parameters $\delta^A_V$, $\delta^A_{NS}$ and $\delta^A_\pm$
for various nucler targets
we performed a global fit on structure function ratio data taken by the EMC\cite{EMC-CCuSn, EMC-CCa1, EMC-CCa2, EMC-Cu}, NMC\cite{NMC-CCaovLi, NMC-HeCCaCCaovLi, NMC-LiC, NMC-BeAlCaFeSnPbovC, NMC-SnovC},
SLAC\cite{SLAC-Fe}, BCDMS\cite{BCDMS-NFe, BCDMS-Fe}, E665\cite{E665-CCaPb, E665-Xe}, JLab\cite{JLab-HeCBe} and CLAS\cite{CLAS-CAlFePb} Collaborations.
It is important to point out that
we impose some kinematical cuts on the experimental data to ensure that
the data points are in the deep inelastic region, namely, $Q^2 \geq 1$~GeV$^2$ and $W^2 \geq 4$~GeV$^2$.
During the fit, we suggest that $\delta^A_V = \delta^A_{NS}$.
Following\cite{KL2025}, we set $\Lambda_{\rm QCD}^{(4)} = 118$~MeV, that
corresponds to world averaged $\alpha_s(M^2_Z) = 0.1180$\cite{PDG}.
We apply 'frozen' treatment of the QCD
coupling in the infrared region (see, for example\cite{frozen-aQCD, gDAS-PDFs-smallx} and references therein), where
$\alpha_s(\mu^2) \to \alpha_s(\mu^2 + M_\rho^2)$ with $M_\rho \sim 1$~GeV.
Such treatment leads to a good description of the data on proton structure function $F_2(x,Q^2)$\cite{KL2025}.

\begin{table}
	\label{table1}
	\begin{center}
		\begin{tabular}{c c c c c c c}
			\hline
			\hline
			& & & & & & \\
			$A$ & $\delta^A_{+}$ & $\delta^A_-$ & $\delta^A_{NS}$ & Targets & $N$ & $\chi^2/n.d.f.$ \\
			& & & & & & \\
			\hline
			& & & & & &\\
			$^4$He & $-0.023 \pm 0.004$ & $-0.017 \pm 0.002$ & $0.17 \pm 0.02$ & $^4$He/$^2$D & $42$ & $0.63$ \\
			$^6$Li & $-0.017 \pm 0.005$ & $-0.014 \pm 0.003$ & $0.29 \pm 0.27$ &  $^6$Li/$^2$D & $17$ & $1.39$ \\
			$^6$Li & $-0.015 \pm 0.005$ & $-0.012 \pm 0.002$ & $0.11 \pm 0.12$ &  $^{12}$C/$^6$Li & $24$ & $1.18$ \\
			$^6$Li & $-0.016 \pm 0.006$ & $-0.018 \pm 0.002$ & $0.27 \pm 0.15$ &  $^{40}$Ca/$^6$Li & $24$ & $1.20$ \\
			$^9$Be & $-0.030 \pm 0.005$ & $-0.018 \pm 0.003$ & $0.40 \pm 0.11$ &  $^{9}$Be/$^{12}$C & $15$ & $0.22$ \\
			$^{12}$C & $-0.044 \pm 0.006$ & $-0.026 \pm 0.002$ & $0.33 \pm 0.04$ &  $^{12}$C/$^{2}$D & $47$ & $1.61$ \\
			$^{27}$Al & $-0.073 \pm 0.008$ & $-0.045 \pm 0.004$ & $0.81 \pm 0.32$ &  $^{27}$Al/$^{12}$C & $15$ & $0.48$ \\
			$^{40}$Ca & $-0.074 \pm 0.007$ & $-0.045 \pm 0.003$ & $0.68 \pm 0.10$ &  $^{40}$Ca/$^{2}$D & $32$ & $2.05$ \\
			$^{56}$Fe & $-0.089 \pm 0.011$ & $-0.057 \pm 0.006$ & $1.26 \pm 0.47$ &  $^{56}$Fe/$^{12}$C & $15$ & $0.72$ \\
			$^{64}$Cu & $-0.101 \pm 0.017$ & $-0.068 \pm 0.009$ & $0.91 \pm 0.35$ &  $^{64}$Cu/$^{2}$D & $10$ & $0.82$ \\
			$^{118}$Sn & $-0.100 \pm 0.004$ & $-0.060 \pm 0.002$ & $0.65 \pm 0.08$ &  $^{118}$Sn/$^{12}$C & $161$ & $0.74$\\
			$^{208}$Pb & $-0.104 \pm 0.025$ & $-0.032 \pm 0.013$ & $0.99 \pm 0.32$ &  $^{208}$Pb/$^{2}$D & $19$ & $0.73$\\
			$^{208}$Pb & $-0.124 \pm 0.012$ & $-0.074 \pm 0.006$ & $2.70 \pm 1.07$ &  $^{208}$Pb/$^{12}$C & $13$ & $0.22$\\
			& & & & & &\\
			\hline
			\hline
		\end{tabular}
	\end{center}
	\caption{The $\delta^A_\pm$ and $\delta^A_{NS}$ parameters extracted from the EMC\cite{EMC-CCuSn, EMC-CCa1, EMC-CCa2, EMC-Cu},
		NMC\cite{NMC-CCaovLi, NMC-HeCCaCCaovLi, NMC-LiC, NMC-BeAlCaFeSnPbovC, NMC-SnovC}, SLAC\cite{SLAC-Fe}, BCDMS\cite{BCDMS-NFe, BCDMS-Fe},
		E665\cite{E665-CCaPb, E665-Xe}, JLab\cite{JLab-HeCBe} and CLAS\cite{CLAS-CAlFePb} data.}
	\label{tbl:parameters1}
\end{table}

The results of our fits for various nuclei are collected in Table~\ref{tbl:parameters1}.
The measured nuclear targets involved into the analysis,
number or data points $N$ and goodness, $\chi^2/n.d.f.$, are presented also.
The fitting procedure was done using the
algorithm as implemented in the commonly used \textsc{gnuplot} package\cite{GnuPlot}.
One can see that reasonably good values of $\chi^2/n.d.f.$ are achieved in most of the cases.
The newly fitted values of $\delta^A_+$, responsible for nuclear shadowing effects, are very close to the ones obtained earlier\cite{RescalingModel-Apps2},
where the analysis has been done for $^4$He, $^{12}$C and $^{40}$Ca only.
It is because $\delta^A_+$ fits are mainly sensitive to the low-$x$ data,
where the same PDFs asymptotics were used in the both calculations.
However, our %present
results for $\delta^A_-$ and $\delta^A_{NS}$, essential for larger $x$
and responsible for description of the EMC and/or anti-shadowing effects, differ from
earlier analyses\cite{RescalingModel-Apps1, RescalingModel-Apps2}.
The main sourses of this difference are
related with several points.
Firstly, present consideration is based on the updated PDFs, which are more accurately
determined in a wide region of $x$ and $Q^2$.
Secondly, here we took into account effects of Fermi motion
not considered previously. Thirdly, we perform the fit in a whole $x$ range,
while analyses\cite{RescalingModel-Apps1, RescalingModel-Apps2} were done in restricted kinematical regions ($x < 0.1$ and $x < 0.7$, respectively).
Moreover, in fact, both these earlier fits were done with $\delta_{NS}^A$ corresponding to study\cite{RescalingModel-1}.
In contrast, here we extract $\delta_{NS}^A$ values independently
%from other considerations
using an extremely extended data set.
Nevertheless, the strong difference in determined $\delta^A_-$ and $\delta^A_{NS}$
leads to not so different results for nuclear modification factor $R^A$ (see below).

\begin{figure}
	\begin{center}
		\includegraphics[width=4.7cm]{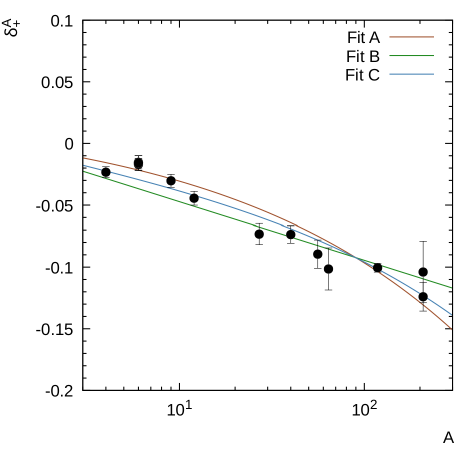}
		\includegraphics[width=4.7cm]{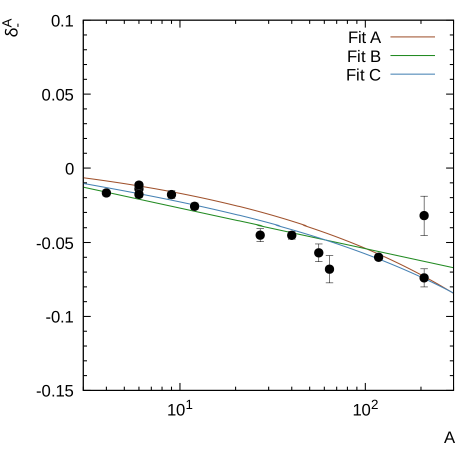}
		\includegraphics[width=4.7cm]{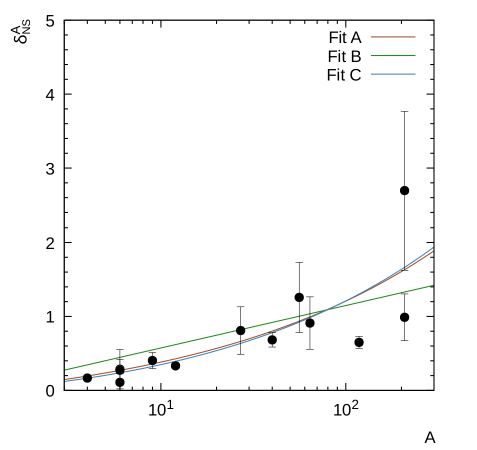}
		\caption{The rescaling parameters $\delta^A_\pm$ and $\delta^A_{NS}$ listed in Table~\ref{tbl:parameters1} and
			fitted in accorgding to (\ref{eq:fitA}), (\ref{eq:fitB}) and (\ref{eq:fitC}) as a function of mass number $A$.}
		\label{fig:1}
	\end{center}
\end{figure}

\begin{table}
	\label{table1}
	\begin{center}
		\begin{tabular}{c c c c c c c}
			\hline
			\hline
			& & & & & & \\
			& $a_+$ & $a_-$ & $a_{NS}$ & $b_+$ & $b_-$ & $b_{NS}$ \\
			& & & & & & \\
			\hline
			& & & & & &\\
			Fit A & $-0.0265$ & $-0.0148$ & $0.3314$ &  &  & \\
			Fit B & $-0.0205$ & $-0.0118$ & $0.2492$ &  &  & \\
			Fit C & $-0.0201$ & $-0.0124$ & $0.3587$ & $0.0287$ & $0.0160$ & $0.1228$ \\
			& & & & & &\\
			\hline
			\hline
		\end{tabular}
	\end{center}
	\caption{The parameters $a_i$ and $b_i$ involved into the fits of the nuclear dependence of $\delta^A_\pm$ and $\delta^A_{NS}$
		according to (\ref{eq:fitA}), (\ref{eq:fitB}) and (\ref{eq:fitC}).}
	\label{tbl:parameters2}
\end{table}

\begin{figure}
	\begin{center}
		\includegraphics[width=4.7cm]{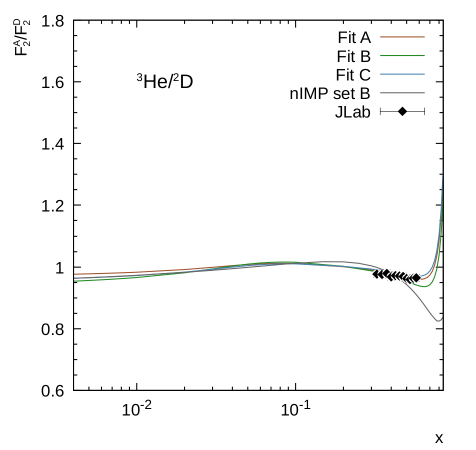}
		\includegraphics[width=4.7cm]{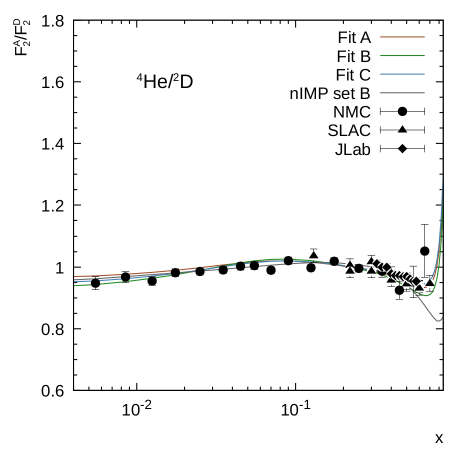}
		\includegraphics[width=4.7cm]{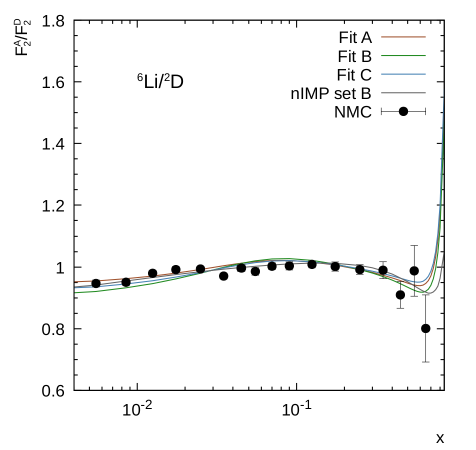}
		\includegraphics[width=4.7cm]{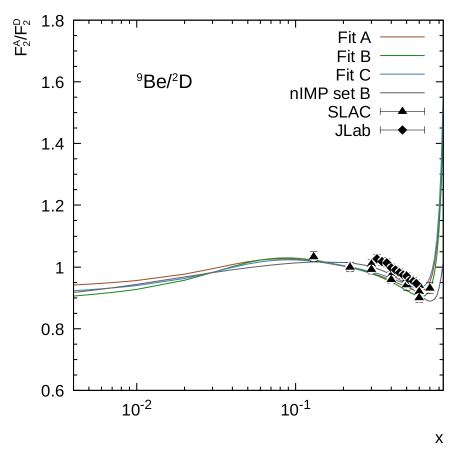}
		\includegraphics[width=4.7cm]{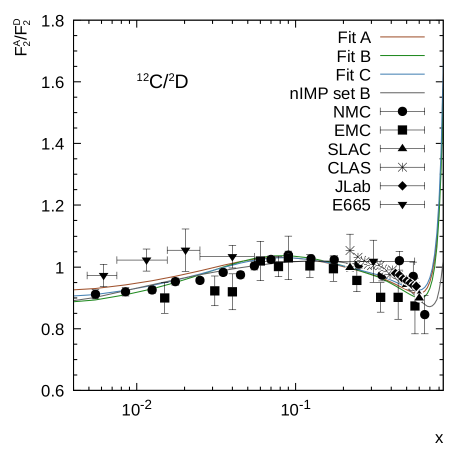}
		\includegraphics[width=4.7cm]{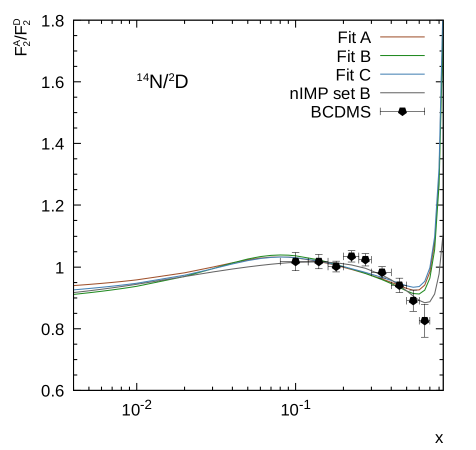}
		\includegraphics[width=4.7cm]{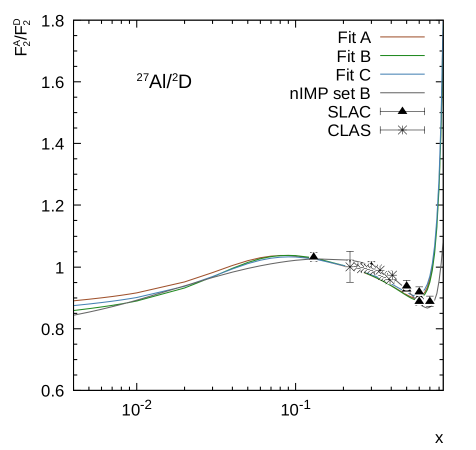}
		\includegraphics[width=4.7cm]{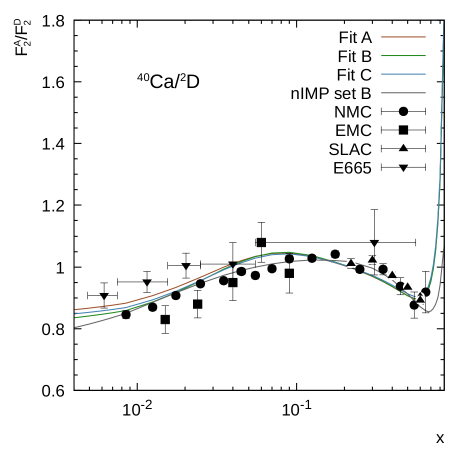}
		\includegraphics[width=4.7cm]{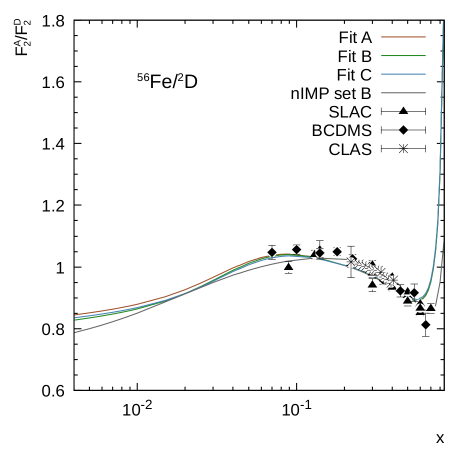}
		\includegraphics[width=4.7cm]{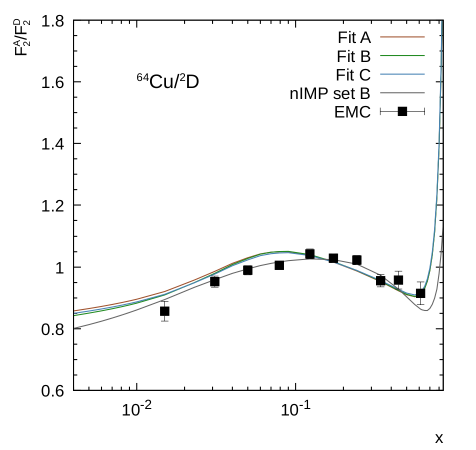}
		\includegraphics[width=4.7cm]{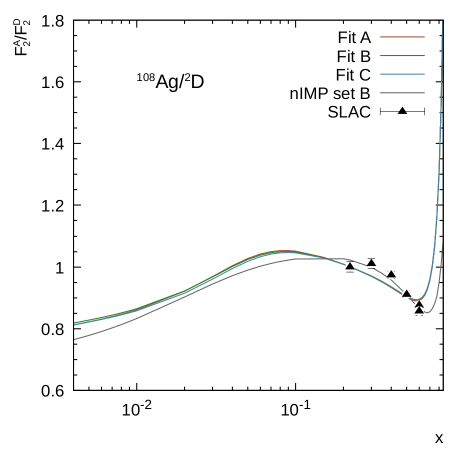}
		\includegraphics[width=4.7cm]{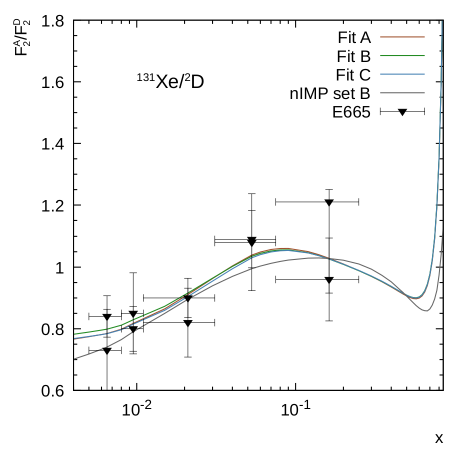}
		\includegraphics[width=4.7cm]{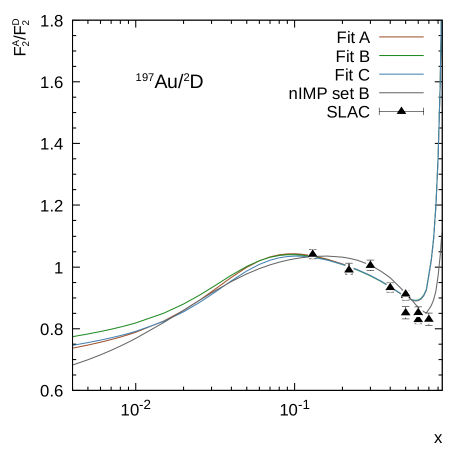}
		\includegraphics[width=4.7cm]{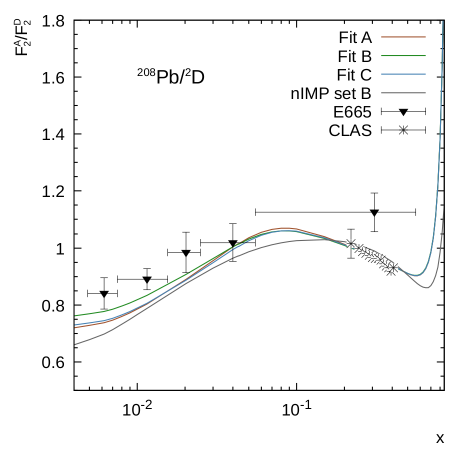}
		\caption{The global fit results of structure function ratios
			$F_2^{A}(x, Q^2)/F_2^D(x, Q^2)$ between different nuclei targets $A$ and deutron.
			Experimental data are from
			EMC\cite{EMC-CCuSn, EMC-CCa1, EMC-CCa2, EMC-Cu},
			NMC\cite{NMC-CCaovLi, NMC-HeCCaCCaovLi, NMC-LiC, NMC-BeAlCaFeSnPbovC, NMC-SnovC}, SLAC\cite{SLAC-Fe}, BCDMS\cite{BCDMS-NFe, BCDMS-Fe},
			E665\cite{E665-CCaPb, E665-Xe}, JLab\cite{JLab-HeCBe} and CLAS\cite{CLAS-CAlFePb}.}
		\label{F2AF2D}
	\end{center}
\end{figure}

\begin{figure}
	\begin{center}
		\includegraphics[width=4.7cm]{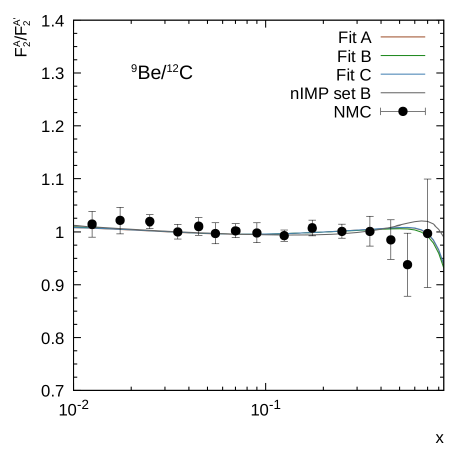}
		\includegraphics[width=4.7cm]{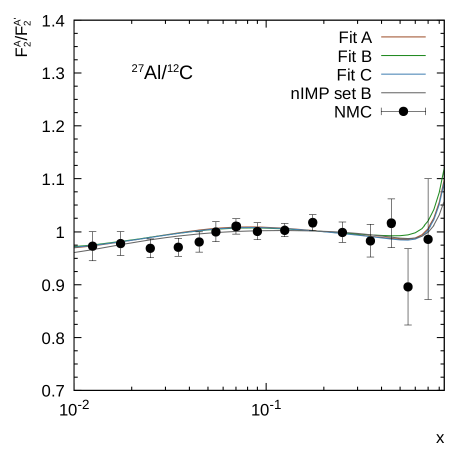}
		\includegraphics[width=4.7cm]{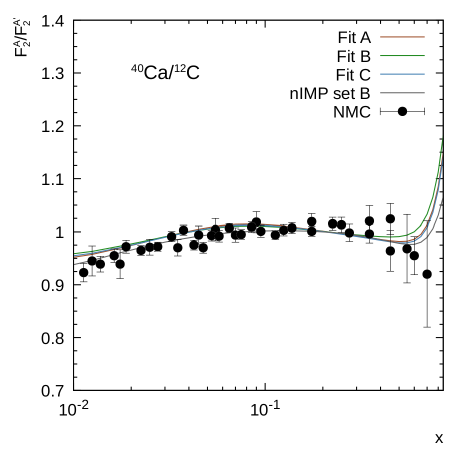}
		\includegraphics[width=4.7cm]{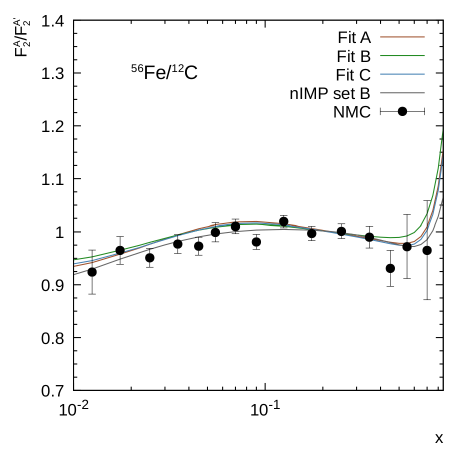}
		\includegraphics[width=4.7cm]{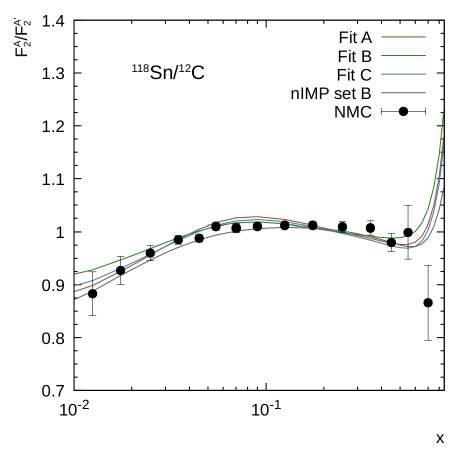}
		\includegraphics[width=4.7cm]{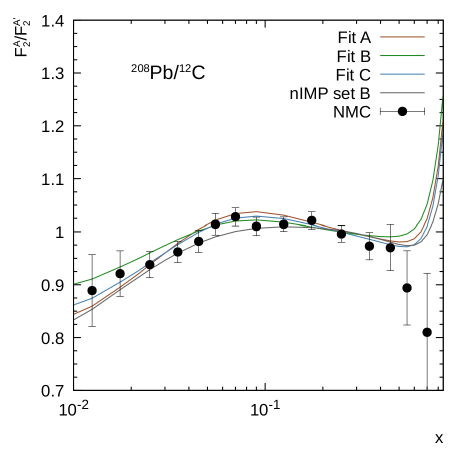}
		\includegraphics[width=4.7cm]{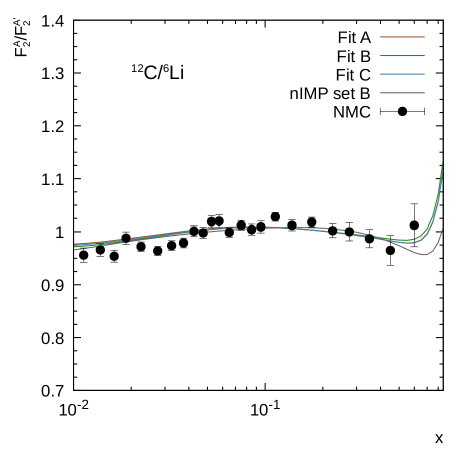}
		\includegraphics[width=4.7cm]{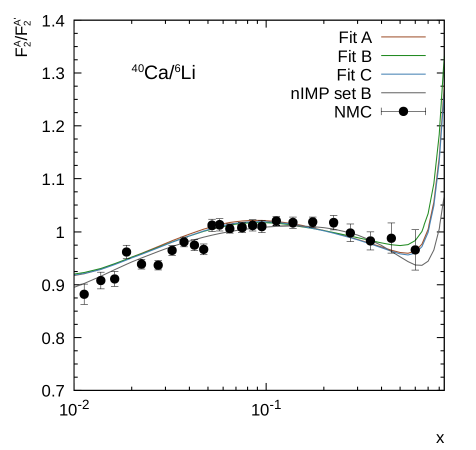}
		\caption{The global fit results of structure function ratios
			$F_2^{A}(x, Q^2)/F_2^{A^\prime}(x, Q^2)$ between different nuclei targets $A$ and $A^\prime$.
			Experimental data are from NMC\cite{NMC-HeCCaCCaovLi, NMC-LiC, NMC-CCaovLi, NMC-BeAlCaFeSnPbovC, NMC-SnovC}.}
		\label{F2AF2A}
	\end{center}
\end{figure}

Now we turn to investigation of the nuclear dependence
of the rescaling parameters $\delta^A_\pm$ and $\delta^A_{NS}$. Such dependence
of nuclear scaling variables is often assumed to be proportional to $A^{1/3}$.
However, there are approaches
where much stronger, proportional to $A^{2/3}$, weaker, proportional to $A^n$ with $n \ll 1/3$,
or even logarithmic $A$-dependencies
\footnote{The logarithmic $A$-dependence appeared, in particular, in the $x$-rescaling procedure \cite{ESegarra21}.}
are favored (see, for example,\cite{QsADependence1, QsADependence2, QsADependence3, QsADependence4, QsADependence5, nCTEQ, ESegarra21} and
the recent review \cite{MKla24} and
references therein).
Here we try to parametrize $A$ dependence of $\delta^A_\pm$ and $\delta^A_{NS}$ in several ways, namely,
\begin{gather}
  \delta_i^A = a_i ( A^{1/3} - 1), \label{eq:fitA} \\
  \delta_i^A = a_i \ln A,  \label{eq:fitB} \\
  \delta_i^A = a_i ( A^{1/3} - 1) + b_i ( A^{-1/3} - 1), \label{eq:fitC}
\end{gather}
\noindent
where $i = \pm$ or $NS$ and we keep in mind that $\delta^A_\pm$ and $\delta^A_{NS}$ have to be zero at $A = 1$.
Hereafter the parametrizations (\ref{eq:fitA}), (\ref{eq:fitB}) and (\ref{eq:fitC}) will be referred as Fit A, Fit B and Fit C, respectively.
Based on the data from Table~\ref{tbl:parameters1}, we have determined all the parameters involved into the Fits
and collect them in Table~\ref{tbl:parameters2}.
The fitted nuclear dependence of $\delta^A_\pm$ and $\delta^A_{NS}$ is shown in Fig.~\ref{fig:1}.
Having this dependence to be
defined, the nuclear medium modification factor can be calculated
for any nucleus $A$, even for unmeasured ones.
It is demonstrated in Fig.~\ref{F2AF2D}, where ratios
$F_2^A(x,Q^2)/F_2^D(x,Q^2)$ for several nuclear
targets are shown in comparison with available deep inelastic data.
Note that some of these
nuclei (for example, $^3$He, $^{14}$N, $^{108}$Ag etc) were not included into our fit due to not sufficient statistics. % of rescaling parameters.
For comparison, we present here the predictions obtained by the nIMP group\cite{nIMP},
which are based on the dynamical parton model
combined with some nuclear models.
We find that predictions obtained with
parametrizations (\ref{eq:fitA}), (\ref{eq:fitB}) and (\ref{eq:fitC}) are
rather close to each other and generally
consistent with the experimental data. % for light as well as heavy nuclei targets.
One can see that both nuclear dependence
and $x$ dependence are more or less reproduced by our calculations.
The strong rise of $F_2^A(x,Q^2)/F_2^D(x,Q^2)$ ratios at $x \geq 0.7$ is related with
Fermi motion taken into account in our analysis.
A reasonably good description of the NMC data
on the ratio of per-nucleon structure functions of one nucleus $A$
to another nucleus $A^\prime$, $F_2^{A}(x, Q^2)/F_2^{A^\prime}(x, Q^2)$, is achieved, as it is shown in Fig.~\ref{F2AF2A}.
The nuclear shadowing effect is decribed well by all the
considered scenarios (\ref{eq:fitA}), (\ref{eq:fitB}) and (\ref{eq:fitC}). %of nuclear depencence of rescaling parameters.
Howewer, Fit A and Fit C as well as nIMP calculations\cite{nIMP} predict a bit stronger shadowing
for heavy nuclei compared to Fit B.
Nevertheless, it is clear that more precision nuclear data,
especially at low $x$, are necessary to distinguish between
different approaches.
Note that for most of the used data the points are taken for
different $Q^2$, so one should consider the lines depicted in Figs. 1 and 2 as interpolations in both $x$ and $Q^2$.

\begin{figure}
	\begin{center}
		\includegraphics[width=4.7cm]{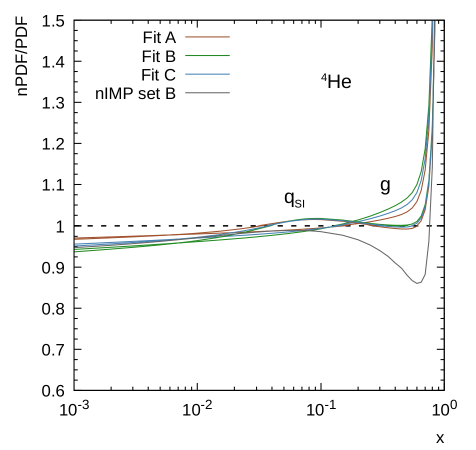}
		\includegraphics[width=4.7cm]{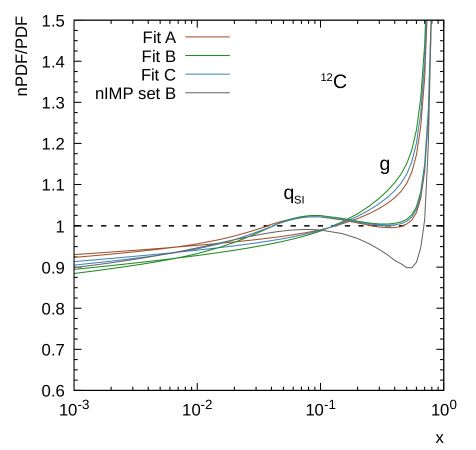}
		\includegraphics[width=4.7cm]{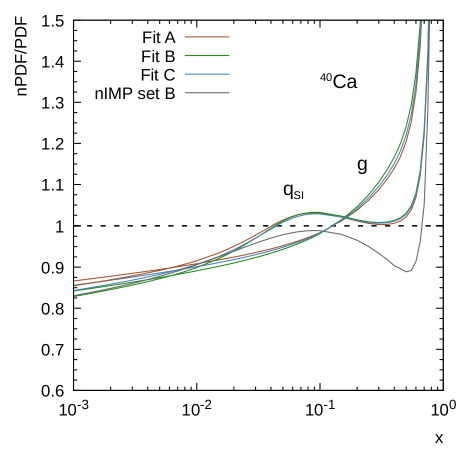}
		\includegraphics[width=4.7cm]{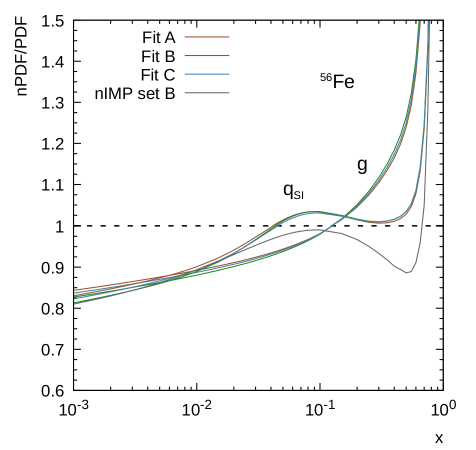}
		\includegraphics[width=4.7cm]{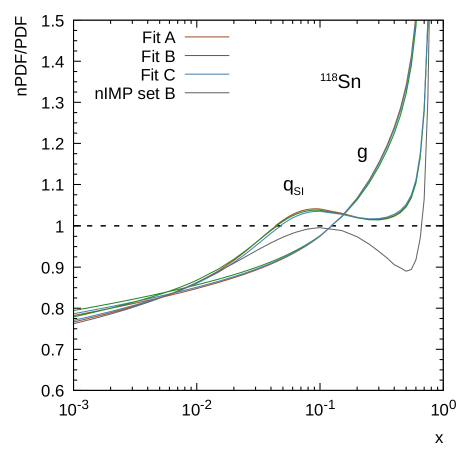}
		\includegraphics[width=4.7cm]{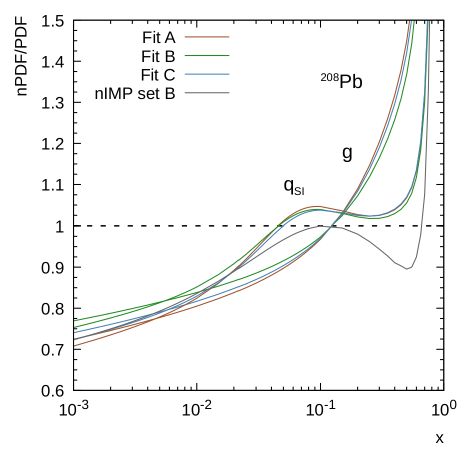}
		\caption{The predicted nuclear modification factors for parton distributions in several
			nuclear targets. %Fixed value $Q^2 = 10$~GeV$^2$ is applied.
			The results
			for gluon nuclear modification predicted by the nIMP group\cite{nIMP} are shown for comparison.}
		\label{fig:3}
	\end{center}
\end{figure}

Next, using the analytical expressions~(\ref{eq1-NSV}) --- (13) for nucleon target, (\ref{va.1a}) --- (\ref{Fermi4}) and~(\ref{Fermi5}) for
nuclear targets and fitted values of rescaling parameters
$\delta^A_\pm$ and $\delta^A_{NS}$ or rather their $A$-dependence, one can give predictions
for nuclear modification of parton distributions.
Our results are shown in Fig.~\ref{fig:3} for several light and heavy nuclei.
As it is well known, nuclear modification is weakly dependent on $Q^2$
%are practically independent on $Q^2$
at low and moderate $Q^2$, so
we show only the results for $Q^2 = 10$~GeV$^2$.
For comparison, we plot also results
for gluon nuclear modification factors predicted by the nIMP group\cite{nIMP}.
We find that the shadowing effect for gluons is, in general, weaker than for quarks,
that is consistent with other studies (see, for example,\cite{nIMP, nCTEQ, nNNPDF3, EPPS21}).
Here, similar to structure function ratios, % discussed above,
Fit A and Fit C predict a bit stronger (weaker) %quark and gluon
shadowing effects
for heavy (light) nuclei compared to Fit B (see Fig.~\ref{fig:1}).
Nevertheless, the difference between these results
are rather small and predicted shadowing effects
are close to the nIMP expectations.
It is in contrast with large-$x$ region,
where we find a significant discrepancy with the nIMP calculations.
In fact, the latter demonstrate a very weak antishadowing at $x \sim 0.1$,
while our calculations
point to the significant antishadowing effects for quarks.
Note that a very big antishadowing of nuclear gluon densities
is predicted by other groups\cite{nNNPDF3, EPPS21}.
Thus, at present it is difficult to draw any
specific inference. We can conclude
again that
precision measurements at future colliders (EiC, EiCC)
are needed to clarify this point.
We hope that this analysis may simplify future
comparisons between the experimental data and theory and could
be helpful in distinguishing different approaches.

\section{Conclusion} \indent
%We presented a brief overview of modifications of parton densities in the nuclear environment based on the rescaling model.

In this work, we
%\textcolor{blue}{
presented a brief overview of the nuclear medium modifications of parton densities based
on the rescaling model. Moreover, we%}
extended the rescaling model to include Fermi motion effects, providing
a consistent description of nuclear
modifications for DIS structure functions and parton distributions
across the whole kinematic range.
Using simple analytical formulas for proton PDFs derived at the LO of QCD coupling earlier,
we have performed a global analysis of available deep inelastic data for different nuclear targets and extract
the corresponding rescaling parameters $\delta^A_\pm$ and $\delta^A_{NS}$.
Then, basing on several assumptions
about their nuclear dependence, we
proposed a simple model to calculate the
parton distributions for any nuclear target, even unmeasured yet.
Finally, the effects of nuclear modifications are investigated
with respect to the mass number $A$.
Our results highlight distinct shadowing and antishadowing behaviors for gluons and quarks,
with implications for future studies of nPDFs with the NLO accuracy.

%\textcolor{blue}{
Note that, according to our current studies (see also\cite{AKotikov89}), in the rescaling model the ratios of the valence and
  non-singlet parts of the parton densities in bound and free nucleons will increase at small values of $x$. This contradicts the results of some
  other studies (see, for example, review\cite{SKulagin16}). Such a discrepancy can be eliminated by studying future experimental data on
  neutrino-nucleon and neutrino-nucleus scattering\cite{MHirai05, KEscola06, RLWorkman22} (see also\cite{MKla24} and discussions therein).%}

The developed approach can be easily used in the
different phenomenological applications.
We hope that it will be also useful for future studies of lepton-nucleus, proton-nucleus and nucleus-nucleus
interactions at modern and future colliders, where nuclear parton
dynamics could be examined directly.

\section*{Acknowledgements} \indent

We thank S.P.~Baranov,
%and
H.~Jung, %\textcolor{blue}{
M.A. Malyshev and N.N. Nikolaev%}
for their interest, very important comments and remarks.
%\textcolor{blue}{
In addition, we express our gratitude to M.A. Malyshev for reading the text of our paper.%}
%  Moreover, we thank M.A. Malyshev also for reading text of our paper.}
This research has been carried out at the expense of the Russian Science Foundation grant No. 25-22-00066, https://rscf.ru/en/project/25-22-00066/.

\bibliography{EMCiNN}

\end{document}